\documentclass[a4paper]{jpconf}
\usepackage{graphicx}
\def\RW{Rolf Wider\o e}

\def\Archiv{Archiv f\"ur Elektrotechnik}
\begin{document}
\title{Birth of colliding beams in Europe, two photon studies  at Adone}
%sBruno Touschek and the road to two photon physics: from betatrons to AdA and  ADONE }
%%Birth of colliding beams in Europe, and the observation of  two-photon collisions  at Adone}
%Preparing a paper using \LaTeXe\ for publication in \jpcs}
\author{L Bonolis$^1$ and G Pancheri$^2$}
%Jacky Mucklow}

\address{$^1$ Max Planck Institut f\"ur Wissenschaftsgeschichte, Boltzmannstra\ss e 22, 14195 Berlin, Germany}
%Production Editor, \jpcs, \iopp, Dirac House, Temple Back, Bristol BS1~6BE, UK}
\address{$^2$ INFN Frascati National Laboratories, Via E. Fermi 40, I00044 Frascati, Italy}

\ead{$^1$ lbonolis@mpiwg-berlin.mpg.de}
\ead{$^2$ pancheri@lnf.infn.it}

%\author{Giulia Pancheri}
%Jacky Mucklow}

%\address{INFN Frascati National Laboratories, Via E. Fermi 40, I00044 Frascati, Italy}
%Production Editor, \jpcs, \iopp, Dirac House, Temple Back, Bristol BS1~6BE, UK}

%jacky.mucklow@iop.org}

\begin{abstract}
This article recalls the birth of the first electron-positron storage ring AdA, and  the construction of the higher energy collider ADONE,  where  early photon-photon collisions were observed. The events which  led the  Austrian physicist Bruno Touschek to propose and   construct AdA will be recalled, { starting with early work on the Wider\o e's betatron during World War II}, up to   the construction of ADONE, and the theoretical contribution to radiative corrections to  electron-positron collisions.
\end{abstract}

\section{Introduction}
Photon-photon physics started its long way towards measurement and observation long time ago. Theoretical calculations of $\gamma \gamma $ processes were developed early before the advent of QED, and then  refined in mid 1950, { as described in other contributions to this session. The first phenomenological studies started with electron-positron collisions, at VEPP-2 and ADONE. The construction of  ADONE has been proposed by Bruno Touschek in 1960 soon after  AdA,  the  first storage ring ever to be built and function, where electron-positron collision were first observed \cite{Bernardini:1964lqa}.

AdA  } opened the way to  higher luminosity, higher energy,  more modern machines and in this contribution { we} shall recall the birth of AdA, and the construction of ADONE, and the { events} which led to the first observation of electron-positron collisions. Since much has been written on this subject, to avoid repetition we shall mostly discuss here less  known parts of the story of AdA, Touschek and ADONE, in particular focusing on the extraordinary combination of events which allowed Touschek to propose AdA in { February} 1960.{  The main reference work for Touschek's life and his contribution to  the construction of both  AdA and ADONE is the biography written by E. Amaldi after Touschek's death in 1978 \cite{Amaldi:1981be}. Further details about  Touschek's life and science can be found in \cite{Bonolis:2011wa,Bernardini:2004rp,Greco:2004np,Bonolis:2005as}.\footnote{ Volume \cite{Greco:2004np} and  the LNF internal Notes since 1953, mentioned later in this article,  are available at http://www.lnf.infn.it/sis/ .}} 
%\\ \bibitem[Amaldi 1981]{amaldi81}Amaldi, Edoardo. 1981. {\it The Bruno Touschek Legacy},  CERN 81-19
%\\ \bibitem[Bonolis 2011]{bonolis2011} Bonolis, Luisa and Pancheri, Giulia. 2011. {\it Bruno Touschek: particle physicist and father of the $ e^+e^-$  collider}, {\it Eur. Phys. J. H} {\bf 36}: 1-61
%\\  \bibitem[Greco 2005]{greco2005} Greco, Mario and Pancheri, Giulia. 2005. {\it Bruno Touschek Memorial Lectures},   Frascati Physics Series, Vol. XXXIII (Frascati 2005), available at http://www.lnf.infn.it/sis/frascatiseries/Volume33/volume33.pdf }

\begin{figure}
\begin{center}
\frame{\includegraphics[width=24pc]{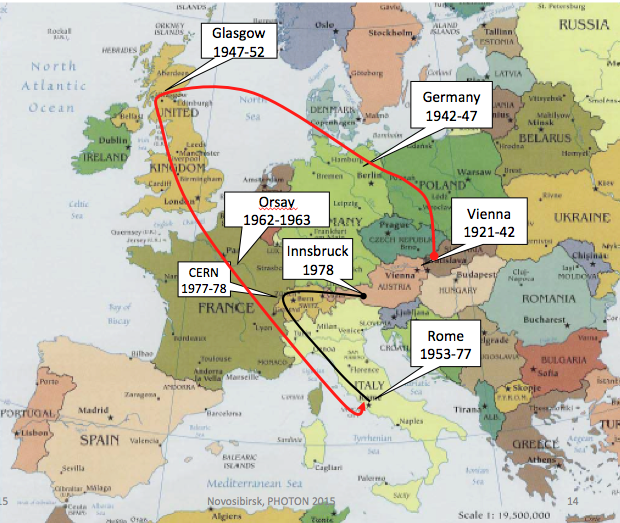}}
\end{center}
\caption{\label{fig:life}A cartoon showing the main stages of Bruno Touschek's life.}
\end{figure}

Touschek's life spans Europe in space and time, from before to after  the Second World War, from the native Austria to Italy, where he gave his greatest contribution to modern day science by proposing the construction of the first electron-positron storage ring.   Fig.~\ref{fig:life} summarizes the major stages of his life. 

The milestones in Touschek's scientific life  can be identified as having taken place in the following   major periods:
\begin{itemize}
\item 1943-45:  during these years, Touschek was in Germany, and worked with Rolf Wider\o e to build a betatron; 
\item { 1947-1952: he was awarded his Ph.D, published several papers on quantum field theory, double $\beta$-decay and meson physics, and was involved in the design and construction of an electron-synchrotron;}
\item 1953-59: these are the years during which Bruno Touschek, who had been given a position in Rome with the newly established  Istituto Italiano di Fisica Nucleare  (INFN, ), deepened his knowledge in theoretical physics, studying the new symmetries and the  breaking of the old ones;
 \item 1960-64: this is the period for which his contribution to modern science is best known, the proposal and construction of AdA, the discovery of the Touschek effect in Orsay, and the first ever observation of $e^+e^-$ collisions;
 \item  November 1960 onwards:  Touschek drafted a proposal for a large collider, ADONE, and  then followed its construction and  developments.
 \end{itemize}

\section{How  Touschek learnt to build accelerators}
\begin{figure}[h]
\resizebox{1\textwidth}{!}{
\frame{\includegraphics[width=13.5pc]{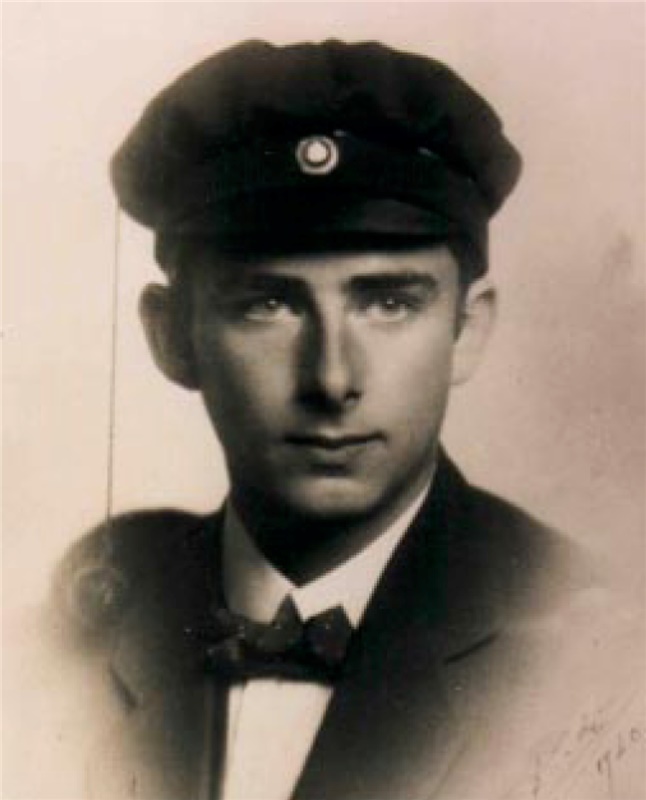}}
\includegraphics[width=25pc]{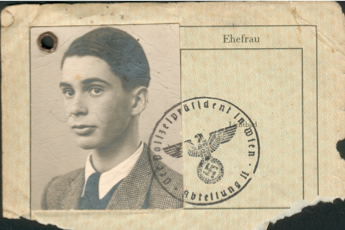}}
\caption{At left, a photograph of the 18 year old \RW \ from \cite{Waloschek:1994qp} . At right,  a photograph of Bruno Touschek from his passport. It  was probably prepared in 1939, for his last travel { to Italy to visit} his maternal aunt Adele, nicknamed Ada. }
\label{fig:BTandRW}
\end{figure}

Bruno Touschek learnt the art of building accelerators in Germany, during World War II, while working on a secret betatron project, led by the Norwegian engineer Rolf Wider\o e \cite{Waloschek:1994qp}, financed by the Aviation Ministry of the Reich, { the  {\it Reichsluftfahrtministerium} (RLM)}. The encounter and collaboration of Touschek and Wider\o e, which ultimately led to the AdA proposal in 1960, follows from a series of rather extraordinary coincidences, which  { originated} on the one side   in Norway, in Trondheim and Oslo,  and on the other  in Vienna and Munich.  We know for sure that in September 1943 Touschek and Wider\o e were already working together on the betatron project, and in fact it is at that time that  Wider\o e mentioned to Touschek  the  idea of oppositely charged colliding particles. But how did they meet? One was a 22 year old  physics student, born in Vienna from a Jewish mother, who died young,  while his  father, who later remarried,  had been an officer in the Austrian army.  The other was an experienced  engineer from { Oslo}, who had done  early work on the theory and construction of accelerators for the  PhD he had obtained in Karslruhe. Not irrelevant to the story is that  Wider\o e's  brother had been working for  the Norwegian underground and was kept prisoner by the Germans.  The desire to help him  is one of the reasons Wider\o e ultimately accepted to come to Germany to build a betatron. Touschek had left Vienna to continue his studies in physics in the relatively anonymity of Hamburg and Berlin, after he had  to stop attending classes at the University, where he had enrolled to study physics, and where he had already shown excellence. In Fig.~\ref{fig:BTandRW}, we show two photographs of Touschek and Wider\o e.

\begin{figure}[h]
\begin{minipage}{20pc}
\includegraphics[width=20pc]{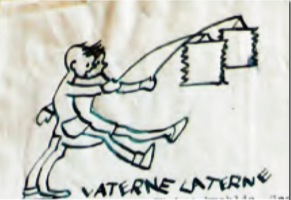}
\centering
\caption{\label{fig:lanterne}A drawing by Bruno Touschek, from a 23 August 1942 letter  from Hamburg.}
\end{minipage}
\hspace{3.5pc}%
\begin{minipage}{14pc}
\centering
\includegraphics[width=6pc]{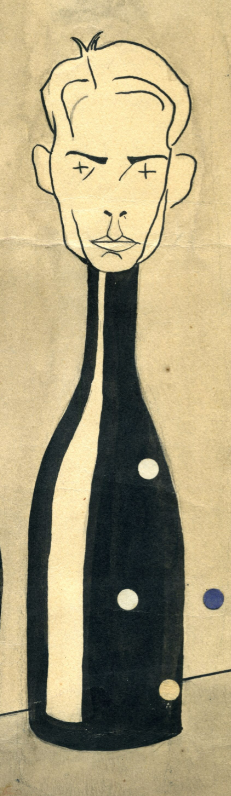}
\caption{\label{fig:bottle}A 1942 portrait of Bruno Touschek, { included in } a letter to his parents.}
\end{minipage} 
\end{figure}

 { We} shall try to separately outline the two stories which came together in Berlin in 1943, and  will start   describing  the path which brought  Touschek  to learn of Wider\o e's work.  % He had moved  to Germany in early 1942, in the hope to be able to continue his studies in physics.
  Bruno  had left Vienna, where he was born in 1921. His  Jewish origin from  mother's side  had brought  many difficulties after the annexation of Austria to Germany in 1938. Enrolled at the University of Vienna to study physics in 1939, at the end of the academic year he had to stop attending classes and could only study at home with books borrowed by his young teacher Paul Urban. Thus he moved to Germany, under the patronage of Arnold Sommerfeld, who had been contacted by Bruno about some small errors in  the fundamental treaty {\it Atombaum und Spektrallinien}, which Bruno had started studying. { In February 1942 } Bruno was in Munich, visiting Sommerfeld, and then went to Hamburg, where he started attending courses at the University, studying and doing odd jobs to make a difficult living. In Figs. ~ \ref{fig:lanterne} and \ref{fig:bottle}, we show two drawings from this period, which Touschek had included in letters to his father and step mother.\footnote{In these letters  Touschek addresses  them  as  {\it Dear parents}.}  
Under Sommerfeld's recommendation, Bruno was unofficially attending classes at both University of Hamburg and University of Berlin, and frequently moved between the two cities. Sometime, probably in fall 1942, he met a girl, also half-jewish, who worked { in Berlin, at Loewe-Opta, a radio and television manufacturing company,} and suggested he could also obtain a job there. And so it happened that Bruno ended up working  with  { Karl A. Egerer, who was at the time the director of a special department within the company, that was now also producing  electronic devices of war interest. Egerer was the editor of the scientific journal {\it \Archiv}, too.} In this journal, in 1928,   \RW \ had published  his article on the theory of betatrons and to this journal, as we shall presently see, he would  send on September 15, 1942 an  article entitled {\it Der Strahlentransformator}{, where he presented the proposal to construct a 15 MeV betatron, and even gave hints about a more powerful 100-MeV machine.} This is the article, accepted but never published, which was to put in motion the RLM  betatron project. We have in fact some evidence from a February 1943 letter
% by Touschek 
to his parents,  that Touschek read this article, { in his capacity as assistant to Egerer,} and  commented upon it to him. In this letter, Touschek tells his parents that Egerer  became excited and started making {\it crazy} plans for some war related project to present to the RLM, or even to Heisenberg.\footnote{ This letter and a translation of it were published in \cite{Bonolis:2011wa}.} Egerer was in contact with  other scientists and engineers gravitating around  various high ranking officers at the RLM, and it is suggestive to think that this is how  the project must have reached them and started its way to realization. Once accepted, the article was classified and could not be published.\footnote{A copy of the article in its proofs was kindly provided to us by Aashild S\o rheim, of the University Radium Museum in Oslo. A more detailed account of the events involving the unpublished article can be found in  \cite{Bonolis:2011wa}.} 
%{\it Bruno Touschek: particle physicist and father of the $ e^+e^-$  collider}, {\it Eur. Phys. J. H} {\bf 36}: 1-61}  
At this point, we  shall now step back  to see the train of events which made  \RW \ submit the article and then come to Hamburg in September 1943  to build the 15 MeV German betatron.  

In the United States, the 1928 article by  Wider\o e  had interested accelerator scientist. His first unsuccessful attempt to build a betatron inspired Ernest Lawrence to build the first cyclotron \cite{Lawrence:1931cb} and was later central  to the construction of the betatron by Donald Kerst, who reported it in  his  {\it Physical Review} articles on the betatron \cite{kerst:1940zz,kerst:1941zz,kerst:1941serber}. It is worth noticing here that the same articles were the subject of the 1941 thesis work by Giorgio Salvini, the physicist who would be called in 1953 to build the Frascati electro-synchrotron and who, in 1960, as Director of the Frascati National Laboratories,   approved the construction of AdA.

The issue of the  {\it Physical Review} with Kerst's article was one of the last to reach Nazi-occupied Norway
and it was read, in the fall of 1941, by the physicist Roald Tangen, from the
Physical Institute of Trondheim University. In Tangen's words [6]: {\it I can well
remember the events of 1941 [\dots.] In the autumn of 1941 the Physics Association
invited me to give a lecture on modern accelerators in Oslo. We had
been denied access to American magazines by then, and we were completely
ignorant of the betatron. A few days before my trip to Oslo a single copy of
the Physical Review arrived in Trondheim by ordinary mail. Mysteriously, it
had found its way to us. It contained an article by Donald Kerst on the first
working betatron. This fitted well in my lecture in which I went on to explain
that Kerst mentioned a German doctorate thesis by a R. Wider\o e in which a
fundamental equation for the betatron was developed. I did not  know anyone
by the name of Wider\o e at the time, but I told my audience that the name
indicated that he could be a Norwegian. As we were to discover soon enough,
Rolf Wider\o e was sitting in the auditorium.} 

Wider\o e had left working on accelerators after his thesis work, and had { joined} the Norwegian branch of Brown-Boveri. But his interest  { was  rekindled  by Kerst's article and he immediately} went back to work to propose a similar machine to be built in Europe. { Thus,  in September 1942 he submitted}  the article to the {\it \Archiv }. And here is where { the two} stories of Touschek and Wider\o e meet. In his autobiography, Wider\o e writes : 
{\it A very strange thing happened when my first article appeared. One day, it
must have been in March or April 1943, several German Air Force officers
came to NEBB [Norsk Elektrisk og Brown Boveri] wanting to speak with me.
Norway had been under occupation since April 1940. I cannot remember
exactly whether there were two or three of them. 
\dots 
%I was standing next to my bicycle because I always cycled to NEBB. 
They asked whether we could go
to the Grand Hotel together to talk about something.
\dots 
% I countered that it was possible, but first I would have to fix my bike. In the Grand Hotel they asked me to return to Berlin with them.
 They said that it could be a matter of some
importance to my brother \dots The German officers hinted that it may be
possible to release my brother if I helped them. This decided things for me,
and I agreed to go to Berlin. Two days later I was flown there for a short
visit, and they told me about their plans to build betatrons.} 
%{\it  I did not hear anything about this \dots until \dots }. 

{ The coincidence of dates, namely Touschek's reading the submitted article, in mid February, and the visit to Wider\o e by the German officers in Spring 1945, suggests that, following Touschek's comments on the article, Egerer   may have contacted the RLM. The interest of the German war authorities   in betatron research and possible war applications is documented in \cite{waloschek2012deathrays} and this may have prompted the contacts with Wider\o e in Oslo. We also notice that Wider\o e had previously tried in vain to obtain a more lenient treatment of his brother's imprisonment conditions \cite{Waloschek:1994qp,Brustad:1998aa}}. Several groups at the time were quite interested in the possibility of having sources of artificially accelerated particles. And even if it was already clear that the betatron could be employed only as a source of X-rays used mainly for medical purposes, building such a machine was of course an exciting  scientific project per se. 

Bruno worked alongside with Wider\o e from 1943 until  March 1945. During this period Touschek contributed with theoretical work, the great part of which was later used for his thesis work at G\"ottingen, at the end of the war. In particular he tackled the problem of the energy losses due to radiation damping, which would define an upper limit for the energy obtainable with the betatron. As the war approached its end,  the betatron, which was almost completed,  was moved from Hamburg to { Kellinghusen},  a supposedly safer location, 40 Km North of Hamburg, in the  property of one of the scientists of the betatron group. As recounted many times, around March 15th, Touschek was arrested and kept in the infamous jail of Fuhlsb\"uttel. He  received some comfort from Wider\o e's 
visits, which brought him books and cigarettes, and promises of release, but   the future destination of prisoners in this jail, mostly Jews,  was  the Kiel concentration camp, some 30 Km North of Hamburg.  And it was to this camp, as   the allied troops were  approaching Hamburg in mid April,  that Touschek was  directed, together with other 200 prisoners, guards in front and guards at the back. Luckily for him and for science,   Touschek  did not reach Kiel: he  fell to the ground, being sick and burdened by  a load of books and few belongings, was shot and left for dead on the side of the road. { Further developments of Touschek's story during the war are outlined in \cite{Bonolis:2011wa}, where the two letters written  by Touschek to his family in June  and October  1945 are presented.}

\section{The birth of $e^+e^-$ colliders}

The construction of AdA  and its final success is the work of many  scientists. Three, among them,   had a pivotal role: Bruno Touschek, who proposed it and contributed to its commissioning, { Giorgio} Salvini who made the Frascati synchrotron to work in 1959 and created, from nothing,  a pool of scientists, technicians and engineers of world capacities, and Edoardo Amaldi, Fermi's youngest collaborator in Rome before the war.
Edoardo Amaldi, who was one of the leading actors in the resurgence of physics in Italy and in Europe after the war,  called to Rome both Touschek and Salvini, two scientists who had, in their training, the knowledge and the mindset of constructing accelerators.  

\begin{figure}[h]
\begin{center}
\frame{\includegraphics[width=24pc]{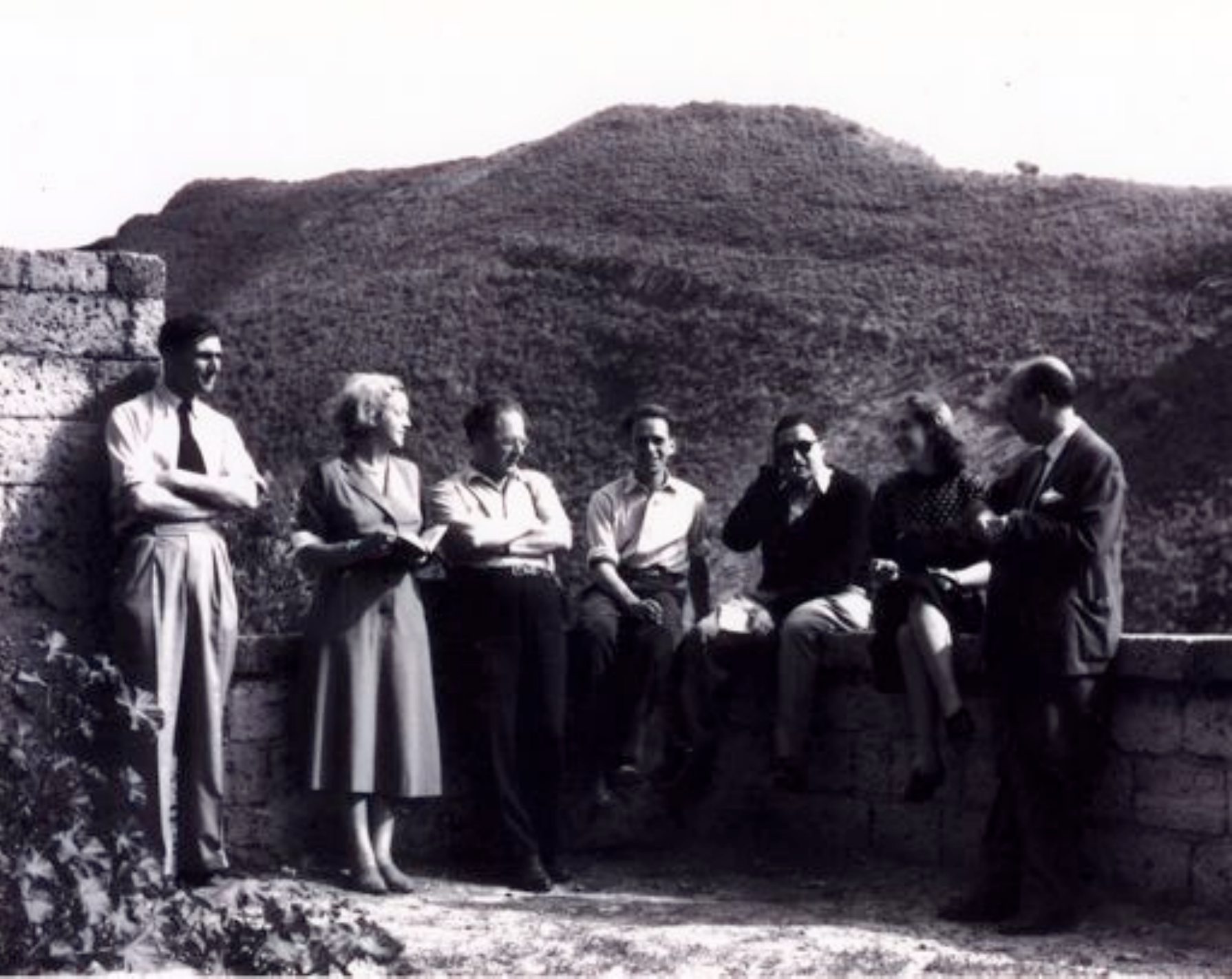}}
\end{center}
\caption{\label{fig:touschek1953}Bruno Touschek (center), in Italy in 1953, at Tuscolo hills with Edoardo and Ginestra Amaldi to his right.}
\end{figure}

{ In Fig.~\ref{fig:touschek1953} we show Touschek with Amaldi and other scientist friends, during a excursion to the Tuscolo hills, above the town of Frascati.}
During his  earlier period in Rome, where he joined as  INFN researcher in 1952,  Touschek does not  seem to have been interested to the synchrotron work. He was keen in expanding his knowledge of theoretical physics and engaged in the challenges posed by the renewed post-war activities and  quantum field theory  formulations, as testified, notably, by his intense correspondence with Wolfgang Pauli during the 1950s. But as the synchrotron approached its completion, his interest grew, and in 1959, he was coming regularly to the Laboratory, attending seminars and meetings. It was one such seminar, by the  Director of the High Energy Physics Laboratory of Stanford University, Wolfgang Panofsky,   which seems to have ignited the spark which  started AdA. { A first reconstruction of this event can be found in \cite{Bonolis:2005as}, where the sources of different recollections are discussed.} A possible date for this seminar is October 26th, 1959, since the list of  1959-60 seminars at Frascati Laboratories shows a seminar by Panofsky on that day. 
Nicola Cabibbo, who had graduated with Touschek a few years before, recalled \footnote{Bruno Touschek's life during the period spanning from his arrival in Italy in 1952 to the building of AdA and ADONE is outlined in the docu-film {\it Bruno Touschek and the art of physics}, by  E. Agapito and L. Bonolis, \textcopyright INFN 2003} that after this seminar  and a discussion on the electron-electron tangential ring collider built in the US, 
%a seminar in Frascati , in Fall 1959,   where
 Touschek asked: {\it Why not  using electrons against positrons?} According to Cabibbo, it was thus  during this seminar that the idea of having electrons and positrons circulate and collide within the same ring, was put forward by Bruno. In the months to follow he pursued the idea and started making some calculations for a draft proposal, which he presented on February 17th, 1960 at a meeting called to discuss the future programs of the Laboratory.  
  
\begin{figure}[h]
\begin{center}
\includegraphics[width=20pc]{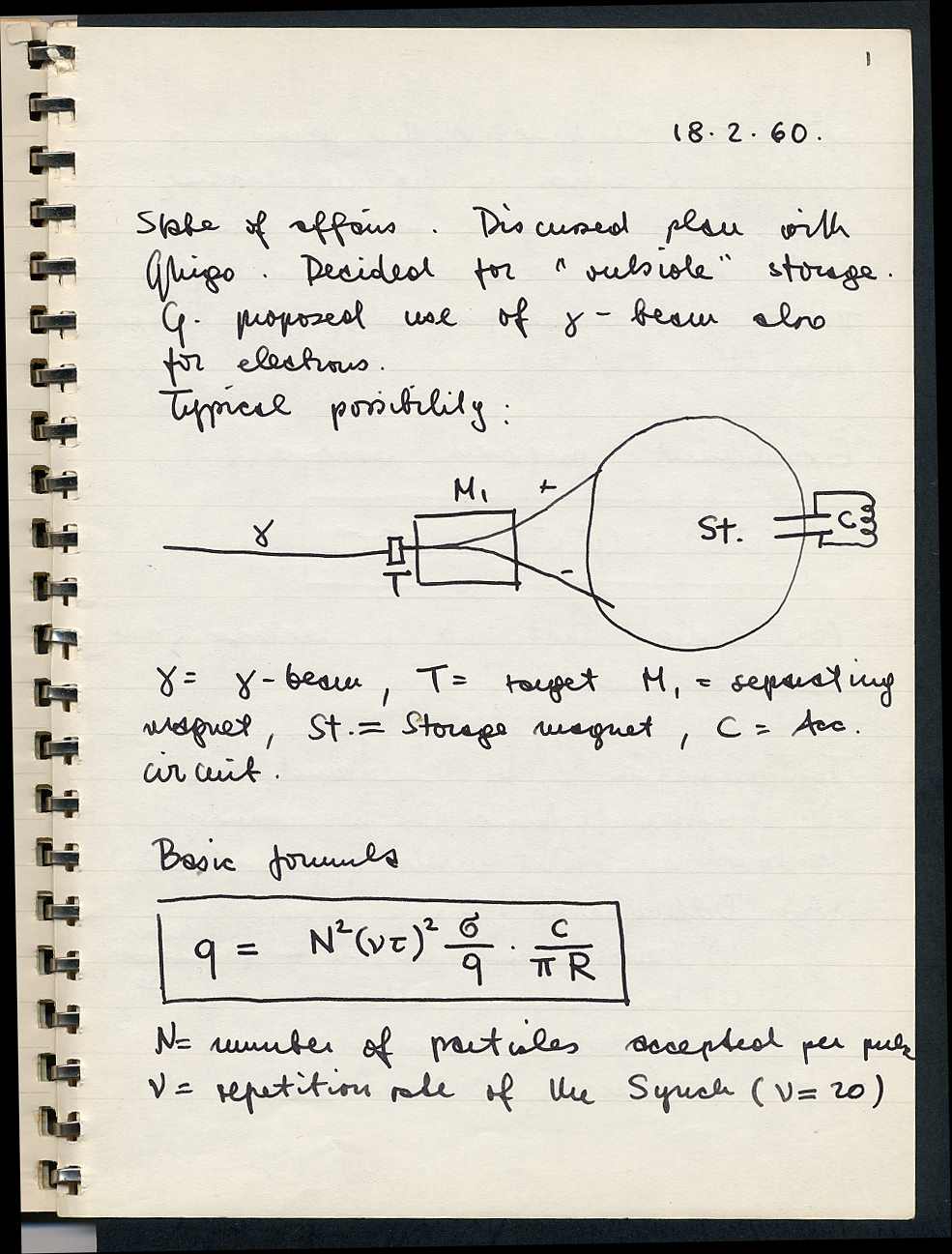}
\end{center}
\caption{\label{fig:quaderno} First page of the notebook where Touschek started the actual   proposal for the construction of AdA. }
% prepared by Bruno Touschek as soon as he was certain that the storage ring principle of AdA was working. Notice the value of the c.m. energy he proposed, namely 3 GeV, chosen so as to observe pairs of all the particles known at the time. }
\end{figure}

 He had envisaged to modify the Frascati electro synchrotron to make it into a storage ring, but this was unthinkable given the expectations of the Frascati physicists to start experimentation with the synchrotron set-up. On February 18th, the day after the meeting, he then started work on a  realistic proposal, { whose  first page  is shown in  Fig. ~\ref{fig:quaderno}.}
 { As a comment to the intense work, which  started taking place in those months, such as the comprehensive calculations  of expected processes \cite{Cabibbo:1961sz},
 we show two (later) drawings by Bruno Touschek, in { Fig.~  \ref{fig:disegni}.} 
 %and \ref{fig:pensatore}.}
  \begin{figure}[h]
%\resizebox{1\textwidth}{!}{
\frame{\includegraphics[width=12pc]{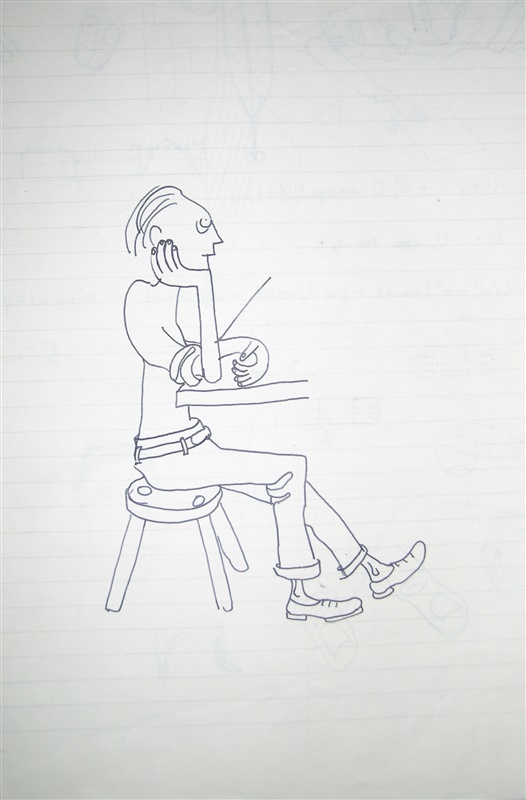}}
\hspace{1.5cm}
\frame{ \includegraphics[width=22pc]{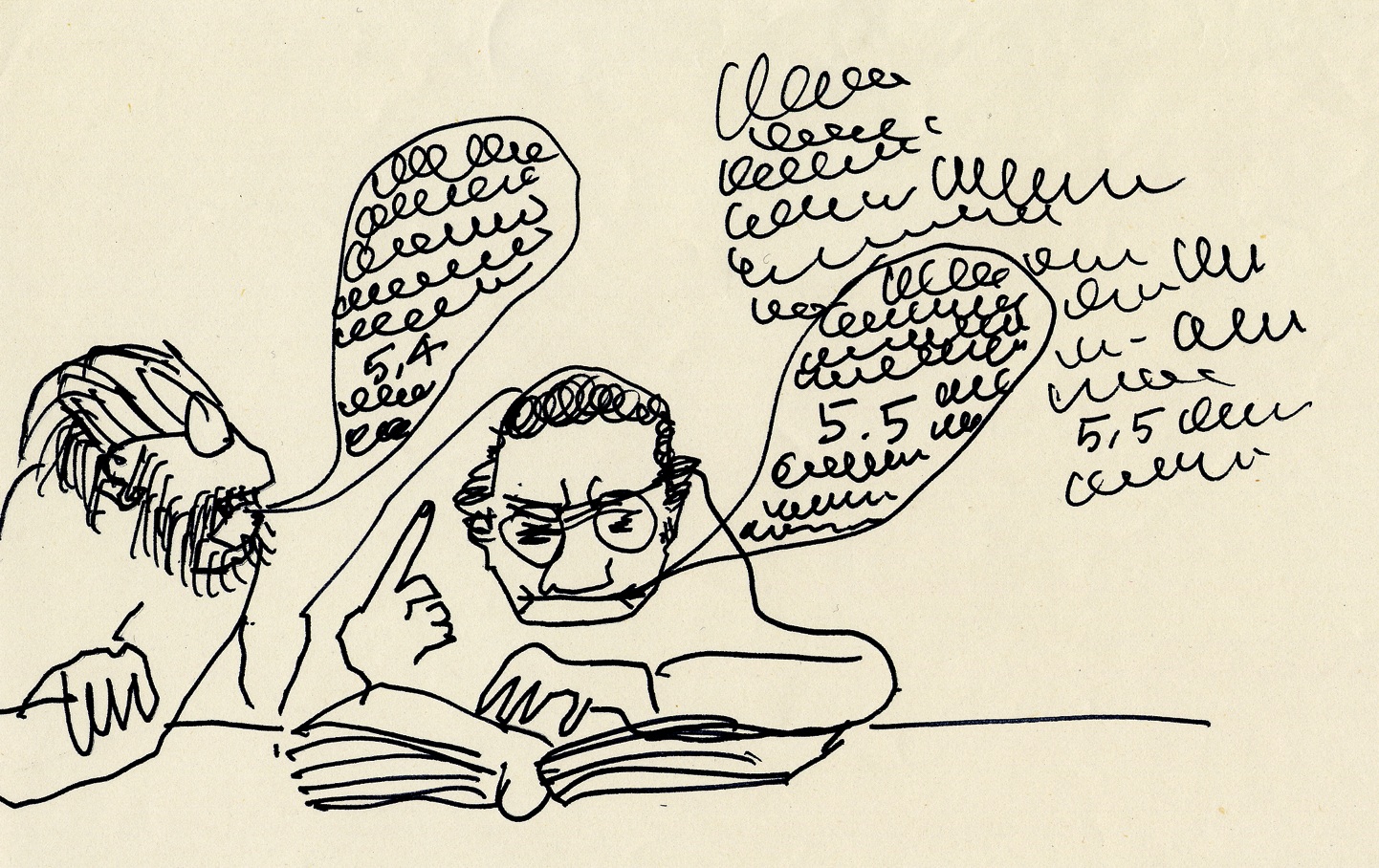}}
%pensatoreDSCN0833}}
\caption{Two drawings by Bruno Touschek. }
%At left, a photograph of the 18 year old \RW \ from \cite{Waloschek:1994qp} . At right,  a photograph of Bruno Touschek from his passport. It  was probably prepared in 1939, for his last travel { to Italy to visit} his maternal aunt Adele, nicknamed Ada. }
\label{fig:disegni}
\end{figure}

{   The proposal  was    approved on March 7th and in less than one year AdA  started functioning \cite{Bernardini:1960nc,Bernardini:1962ncc}.} 
As Pierre Marin would later recall \cite{Marin:2009}, a team of first class scientists designed and  built a small ring in which they made  electrons and positrons circulate, in opposite directions, for hours. 
They were Carlo Bernardini,  Gianfranco Corazza, Giorgio Ghigo, Mario Puglisi, Ruggero Querzoli, Giancarlo Sacerdoti, Peppino di Giugno and, of course Bruno Touschek.  AdA started working in February 1961 but the road to prove the viability of this type of accelerator  was long. It took two yeas and the transfer of AdA to Orsay before obtaining proof that annihilation had taken place.\footnote{The transfer of  AdA to Orsay is described in the docu-film {\it Touschek with AdA in Orsay}, by E. Agapito, L. Bonolis and G. Pancheri, \textcopyright  INFN 2013.}

\section{Photons with AdA and  two photon observations with ADONE}
AdA's luminosity had been too low to observe annihilation into final particles. This had been true in Frascati, but remained true even after AdA had been transported to Orsay to make use of  the higher intensity of the electron beam from  the linear accelerator, jokingly called by students and researcher's alike {\it OLGA}, Orsay Linear Great Accelerator.  Instead,  the proof of collisions had come from the observation of events consistent with single bremsstrahung \cite{Altarelli:1964aa}, namely
\begin{equation}
e^+e^-\rightarrow e^+e^- \gamma \label{eq:brem}
\end{equation}
Touschek however had been firmly convinced that such type of machine would work, and, as early as November 1960, less than a year after he had proposed to build AdA, a proposal to build a much bigger and more powerful machine was put in writing and presented to the Frascati Laboratory management. In Fig. ~\ref{fig:adoneproposal-Permanifesto} we show the typewritten draft, which was transformed in a joint internal laboratory note \cite{Amman:1961ad} a few months later. 
\begin{figure}
%old\begin{center}
%Éto put back\vspace{1cm}
$\vcenter{\hbox{
\frame{\includegraphics[width=20pc]{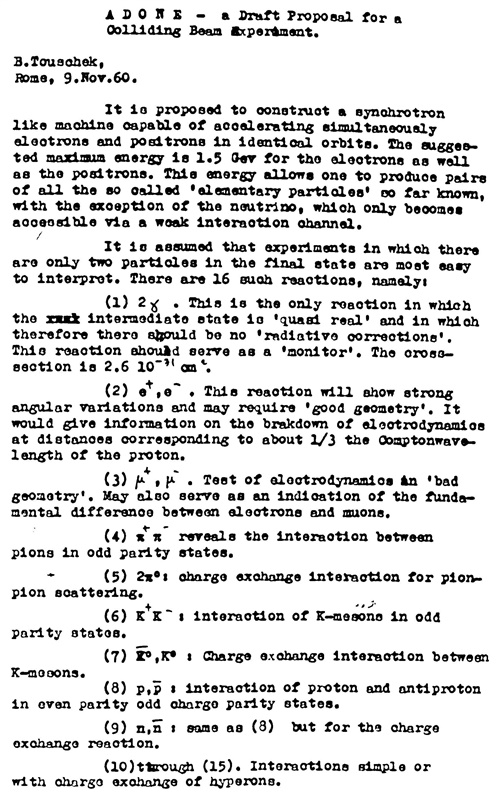}}}}$
%É ti put back \hspace{1cm}
%old\vspace{1cm}
  \hspace*{.2in}
$\vcenter{\hbox{\frame{\includegraphics[width=16pc]{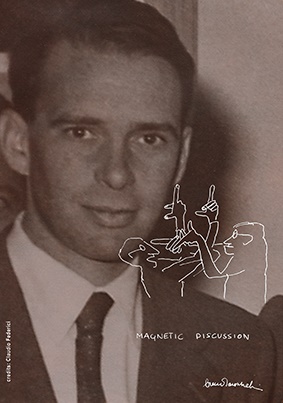}}}}$
%old\end{center}
\caption{\label{fig:adoneproposal-Permanifesto}At left we show the draft of the  proposal for the construction of ADONE prepared by Bruno Touschek as soon as he was certain that the storage ring principle of AdA was working. Notice the value of the c.m. energy he proposed, namely 3 GeV, chosen so as to observe pairs of all the particles known at the time. { At right  Bruno Touschek  in a photograph in the second half of 1950s and one of his drawings, most probably from  the early AdA period,  elaborated by C. Federici. }}
\end{figure} 

\noindent  AdA remained  in Orsay, at the Laboratoire de l'Acc\'el\'erateur  Lin\'eaire, two years, 1962-1964. During this time,   important 
%charge space \cite{Bernardini:1964lqa} {\bf AdA terzo lagoro o secondo?} 
effects, such as the Touschek effect \cite{Bernardini:1997sc}, were discovered. 
  After AdA's final measurements in Orsay in 1964 and the confirmation of collisions through Eq.~(\ref{eq:brem}) \cite{Bernardini:1964lqa}, the Italian team returned to Rome and Frascati, and the construction of ADONE started in earnest. Touschek followed the construction of ADONE, often contributing to  discussions about beam instability  problems, as described in  \cite{Bernardini:2015wja}. He was also particularly worried about radiation problems, whose calculation became the central focus of the theory group  gathered around him in Frascati. 
  { Calculations of radiative effects at ADONE's energies had initiated with the single bremsstrahlung, 
  %Eq. ~(\ref{eq:brem}).
   %{\bf These theoretical calculation 
   and     continued, under Touschek's leadership and guidance, with double bremstrahlung \cite{Greco:1967zz}   and  photon resummation to all orders \cite{Etim:1966zz,Greco:1967zza,Etim:1967aa,Pancheri:1969}.  ADONE started functioning  at the end of  1968 and began taking data in 1969.
 
   Concerning two photon  processes,  early in its operation ADONE observed   the process which had been  proposed by Touschek as monitor for the machine luminosity, i.e.  annihilation into two photons: }
  \begin{equation}
  e^+e^-\rightarrow \gamma \gamma
 \end{equation}
{ This process, which was   observed  in Novosibirsk with VEPP-2 \cite{Balakin:1971hz} and in Frascati, with ADONE \cite{Bacci:1971nk} in the range   with $E_{beam}=0.7-1.2\ GeV$,  
was aimed 
%[see at bacci-2gamma-LNF\_71\_016]  
 at verification of QED. Other channels of annihilation into pairs of muons, pions, kaons followed, as Touschek had  envisaged in his draft proposals. 

 Soon, both  ADONE and VEPP-2 reported  the  observation of the more complicate final state $e^+e^-\rightarrow  e^+e^- + X$, such as production of an electron-positron pair accompanied by  hadrons, or by a second  electron-positron pair, or  a muon pair. As recalled in the contribution to this conference by  Elena Pakhtusova, the first observations of the process }
  \begin{equation}
 e^+e^-\rightarrow e^+e^- \gamma \gamma \rightarrow  e^+e^-   e^+e^-  
 \end{equation}
{ were  made  with VEPP-2  \cite{Balakin:1971ip}
%V.E. Balakin, A.D. Bukin, E.V. Pakhtusova, V.A. Sidorov, A.G. Khabakhpashev,
%Phys.Lett. B34 (1971) 663-664}} 
and with ADONE \cite{Bacci:1972tk}. The ADONE results had also been earlier  reported  at  the 1971  Bologna Conference in April,\footnote{  International Conference on Meson Resonances and Related Electromagnetic Phenomena, Bologna 1971.}  and at Cornell,\footnote{The International Conference on Electrons and Photons Interactions, Cornell  1971.} in August 1971 \cite{Bernardini:1971aa}.
%{ \bf   and published Lettere Nuovo Cimento, vol. 3  1972\cite{2photon1972}.}

Soon, there followed the first observation of the process
\begin{equation}
e^+e^-\rightarrow e^+e^- \gamma \gamma \rightarrow  e^+e^-   \mu^+\mu^-  
\end{equation} 
{ which was reported in \cite{Barbiellini:1973zp}.
%Barbiellini et al, PRL1974 (submitted dec 1973) LNF\_74\_009,  at Barbiellini et al, PRL1974 (submitted dec 1973)  at $\sqrt{s}=2.7 \ GeV$ and was the first observation of this process. 
Later, in 1979, the results  of various other measurements of the two photon collision  channels at ADONE, namely }
\begin{equation}
e^+e^-\rightarrow e^+e^- \ X \ \ \ \ \ \ \ \ \ \  \ X=\gamma \gamma \rightarrow e^+e^-,\   \mu^+\mu^-,\  \pi^+\pi^-,\  \eta'(958)
\end{equation} 
{  which  had been taken   in the energy range  $\sqrt{s}=750=1500\ MeV$, were  reported in \cite{Bacci:1979ac}.
 The complete history, as well as the theoretical developments which led to this field of photon-photon physics, can be found  in other talks in the historical session  of this conference. In particular, the theoretical developments, to which the group in Frascati and Rome contributed as well, are illustrated  in  the talks by I. Ginzburg and F. Kapusta. 
 
 As a final comment to this brief historical contribution, we show in { the right hand panel of { Fig. ~\ref{fig:adoneproposal-Permanifesto} a   photo of Bruno Touschek, with his well known  drawing  entitled {\it Magnetic discussion}.}

\section*{Acknowledgements}
{ G.P. thanks  Valery Telnov and the other organizers, who  have invited her  to present this talk and allowed it to be done remotely. The collaboration of the Communication Service of Frascati National Laboratories is gratefully acknowledged. We are also   very grateful to the Touschek family who holds the copyright of photos and drawings by Bruno Touschek and provided  us with unpublished material from  Touschek's letters home during WWII.  }
\section*{References}
\bibliography{touschek-photon2015}
%photon2015}
\end{document}